# Large adiabatic temperature and magnetic entropy changes in EuTiO$_3$

A. Midya and P. Mandal,

Saha Institute of Nuclear Physics, 1/AF Bidhannagar, Calcutta 700064, India

Km Rubi, Ruofan Chen, Jian-Sheng Wang, and R. Mahendiran[1]

Physics Department, 2 Science Drive 3, National University of Singapore, Singapore 117551, Republic of Singapore

G. Lorusso and M. Evangelisti

Instituto de Ciencia de Materiales de Aragón(ICMA) and Departamento de Fisica de la Materia Condensada, CSIC-Universidad de Zaragoza, 50009, Zaragoza, Spain

**Abstract**

We have investigated the magnetocaloric effect in single and polycrystalline samples of quantum paraelectric EuTiO$_3$ by magnetization and heat capacity measurements. Single crystalline EuTiO$_3$ shows antiferromagnetic ordering due to Eu$^{2+}$ magnetic moments below $T_N$ = 5.6 K. This compound shows a giant magnetocaloric effect around its Neel temperature. The isothermal magnetic entropy change is 49 Jkg$^{-1}$K$^{-1}$, the adiabatic temperature change is 21 K and the refrigeration capacity is 500 JKg$^{-1}$ for a field change of 7 T at $T_N$. The single crystal and polycrystalline samples show similar values of the magnetic entropy change and adiabatic temperature changes. The large magnetocaloric effect is due to suppression of the spin entropy associated with localized 4f moment of Eu$^{2+}$ ions. The giant magnetocaloric effect together with negligible hysteresis, suggest that EuTiO$_3$ could be a potential material for magnetic refrigeration below 20 K.

---

[1] Corresponding author (phyrm@nus.edu.sg)

I. Introduction

Refrigeration has become an essential technology in our modern society. However, the traditional gas compression refrigerators which use ozone depleting volatile refrigerants such as chlorofluorocarbons (CFC), hydrochlorofluorocarbons (HCFC) are reaching its technical boundaries in achieving further improvements.[1,2] Therefore, In addition to further developing the gas compression technology, scientists and engineers have begun to explore alternative environment-friendly and energy-efficient cooling technologies to replace the conventional gaseous refrigeration. Magnetic refrigeration can provide such an opportunity. This technology exploits the magnetocaloric effect (MCE) in solid-state magnetic refrigerant, i.e., the change of the adiabatic temperature ($\Delta T_{ad}$) and isothermal magnetic entropy ($\Delta S_m$) of the functional material upon applying or removing the external magnetic field. The mainstream in this field is to find new materials exhibiting a large MCE close to room temperature for domestic and industrial uses. However, low-temperature refrigeration is important for basic research and some specific technological applications such as space science and liquefaction of hydrogen in fuel industry.[1,3] Although $\Delta S_m$ can be large for first-order phase transition in which magnetic and structural transitions are strongly coupled,[4,5,6] refrigerant capacity of the material may be reduced due to thermal hysteresis and mechanical instability of the compounds after thermal cycles. Hence, materials exhibiting second-order phase transition with negligible hysteresis and large MCE are widely explored.[7]

Materials containing $Eu^{2+}$ ($4f^7$) and $Gd^{3+}$($4f^7$) ions can show a large magnetocaloric effect since these two ions have large total angular momentum ($J = S = 7/2$ and $L = 0$). In this context, $EuTiO_3$ is interesting since it has divalent Eu ion with a large magnetic moment and tetravalent Ti ion with no magnetic moment. $EuTiO_3$, a G-type antiferromagnet below $T_N =$ 5.5±0.2 K[8,9], has attracted much attention in recent years due to the observations of magnetodielectric effect in single crystal[10], strain driven ferromagnetism and ferroelectricity in thin film.[11] It is also unique among the rare earth titanates ($R$TiO$_3$, where $R$ is the rare earth ion) because only the Eu$^{2+}$ ion adopts divalent state instead of the trivalent state adopted by other ions ($R$ = Gd, Y, etc.). Although MCE in $R$TiO$_3$ with $R$ = Dy, Ho, Tm, Yb and Er was reported by Yantao Su *et. al.*[12], it has not been reported in EuTiO$_3$. Recently, Rubi *et al.* reported MCE in Eu$_{1-x}$Ba$_x$TiO$_3$ for the 0.1 ≤ $x$ ≤ 0.9 series.[13] However, the adiabatic temperature change of these compounds as well as MCE of the parent compound, EuTiO$_3$ has not been reported so far. In this regard, we have studied the magnetization and magnetocaloric effect in EuTiO$_3$, both in single crystal and polycrystalline samples. We calculated the magnetization and magnetic entropy change using mean field treatment of the Heisenberg model and fitted with experimental data of single crystal EuTiO$_3$.

**II. Experimental Details**

Polycrystalline EuTiO$_3$ sample was prepared by solid-state reaction method using pre-heated Eu$_2$O$_3$ and TiO$_2$ in reduced atmosphere (5% H$_2$ and 95% Ar). The polycrystalline samples were prepared independently at the National University of Singapore and the Saha Institute of Nuclear Physics (India) and characterized by X-ray diffraction and MCE by magnetization measurements. Stoichiometric amounts of these compounds were mixed properly and pressed into feed and seed rods for single crystal growth. These rods were heated at temperature 1050°C for 30 h in the presence of reduced atmosphere (5% H$_2$ and 95 % Ar). The single crystal was grown by traveling solvent floating zone technique using an image furnace (NEC) under the reduced atmosphere at the Saha Institute of Nuclear Physics, Kolkata, India. The X-ray diffraction pattern of powdered sample of the single crystal reveals that the compound is cubic only at room temperature and single phase. The temperature and field dependent dc magnetization were carried out by a SQUID VSM (Quantum Design). The magnetic field was applied along the [001] axis of the crystal. Heat capacity of single crystal sample down to 2 K was measured by relaxation technique in PPMS (Quantum Design). Heat capacity from 50 K down to 300 mK on polycrystalline sample was also measured in a PPMS, equipped with $^3$He option and installed at the University of Zaragoza. The polycrystalline sample consisted of thin pressed pellets (~1 mg) thermalized with ~0.2 mg of Apiezon N grease, whose contribution was subtracted using a phenomenological expression.

**III. Results and discussion**

Fig. 1(a) shows the temperature dependence of the zero-field-cooled (ZFC) and field-cooled (FC) dc susceptibility $\chi_{dc} = M/H$ measured at $H = 100$ Oe. No significant difference has been observed between ZFC and FC curves. The dc susceptibility increases sharply with decreasing temperature and shows a peak at $T_N = 5.6$ K, which corresponds to AFM ordering of Eu$^{2+}$ moments. To get a clear knowledge about the magnetic interaction, the inverse susceptibility $1/\chi_{dc}$ versus temperature has also been plotted in Fig. 1(a). The linear behavior over a wide range of temperature indicates that it follows the Curie-Weiss (CW) law $\chi = C/(T - \theta_{cw})$. A linear fit to high temperature yields Curie constant ($C = 8.26$ emu K/ mole Oe) and positive Curie-Weiss temperature $\theta_{CW} = 3.1$ K. The value of $\theta_{CW}$ is consistent with the previously reported result.[13] The effective moment estimated from the relation $P_{eff} = \left(\frac{3k_B C}{N}\right)^{1/2}$ is 8.1 $\mu_B$, which is very close to the theoretically expected value of $P_{eff} = 2\sqrt{J(J+1)}\mu_B = 7.94$ $\mu_B$ corresponding to the 4f$^7$ spin configuration of Eu$^{2+}$ ($J = 7/2$). The temperature dependence of the specific heat ($C_p$) for singe crystalline EuTiO$_3$ is displayed in Fig.

1(b). With decreasing temperature, $C_p$ decreases down to 10 K and then increases rapidly and exhibits a very sharp λ-like peak at $T_N$, which is shown in the upper inset of Fig. 1(b). Similar to earlier reports,[10, 11] the cubic to tetragonal structural transition in EuTiO$_3$ is clearly visible as an anomaly in $C_p(T)$ at $T_{st} \simeq 281$ K, which has not been observed in polycrystalline sample.[14] In the temperature range 20-200 K, the zero-field $C_p$ can be fitted well with the combined Debye plus Einstein models.[15]

In order to investigate the influence of applied magnetic field on the magnetic ground state, we have measured the field dependence of magnetization (*M*) in the neighborhood of the magnetic transition and beyond. Some representative plots of the field dependence of magnetization for EuTiO$_3$ are presented in Fig. 2. A qualitative similar behavior has been observed in polycrystalline Eu$_{1-x}$Ba$_x$TiO$_3$ for $0.1 \leq x \leq 0.9$ compounds.[13] Below $T_N$, isotherms vary almost linearly with the field in the low-field region and the slope changes as the magnetic field exceeds a critical field value ($H_c$) and at higher fields *M* tends to saturate. The critical field $H_C$ refers to the critical field for the spin-flop transition. Above $T_N$, *M* smoothly increases with magnetic field and a Brillouin-like behavior has been observed. At $T = 2$ K and $\mu_0 H = 7$ T, the observed value of *M* is 7.0 $\mu_B$ per formula unit for EuTiO$_3$, which exactly matches with $M_S = g\mu_B S = 7\mu_B$, expected for complete alignment of all Eu$^{2+}$ spins. Inset of Fig. 2(a) displays *M(H)* measured at $T = 2$ K while sweeping the field in the direction $0 \rightarrow +7$ T and $+7$ T$\rightarrow -7$ T$\rightarrow +7$ T. No hysteresis is observed even at low magnetic fields. Von Ranke *et al.* calculated magnetization and magnetic entropy change in EuTiO$_3$ using mean field treatment of the Heisenberg model.[16] We followed their approach and calculated the magnetization isotherms at selected temperatures using equation (13) in reference (16). We plot the experimental and calculated data together in Fig. 2(b). We used $J_1/k_B = -0.0407$ K, and $J_2/k_B = 0.0828$ K, $g = 2$, where $J_1$ and $J_2$ are the nearest neighbors and next nearest neighbors exchange interactions, $g$ is the gyromagnetic ratio, respectively. The calculated data (solid line) and experimental data (symbol) match perfectly for all temperatures and magnetic field except at $T = 5$ K near field 1 T. These remarkable accurate fittings are achieved by adjusting $J_1$ and $J_2$ values 10 % and 20 % higher, respectively than the values used by Katsufuji and Takagi.[10] The reason for this is that Katsufuji and Takagi used mean-field theory predictions of $T_N$ and $\theta_{CW}$ to determine uniquely $J_1$ and $J_2$. However, mean-field theory typically over-estimates the value of $T_N$.

We estimate the isothermal magnetic entropy change as a function of temperature for the field variation from 0 to 7 T, using the Maxwell relation $-\Delta S_m(T, H) = \sum_i \frac{M_{i+1} - M_i}{T_{i+1} - T_i} \Delta H_i$ , where $M_i$ and $M_{i+1}$ are the experimentally measured values of magnetization for a magnetic field $H_i$ at temperatures $T_i$ and $T_{i+1}$ , respectively. Thermal variation of the magnetic entropy change $\Delta S_m$ for different magnetic fields are shown in Fig. 3 (a) with theoretical estimates of $\Delta S_m$ from mean field model using equation (14) in reference (16). The curves present a characteristic shape with a maximum near the antiferromagnetic ordering temperature $T_N$. While the calculated data using the mean-field treatment exactly follow the experimental data in paramagnetic region, a slight deviation has been observed around $T_N$. The sign of $\Delta S_m$ is negative down to the lowest temperature and the maximum value of $\Delta S_m$ increases with field reaching a value 49 J kg$^{-1}$K$^{-1}$ for a field change of 7 T. The adiabatic temperature change $\Delta T_{ad}$ is the most significant parameter of the magnetocaloric effect and it is the isentropic temperature difference between $S(0, T)$ and $S(H, T)$. $\Delta T_{ad}$ may be calculated from the field dependent magnetization and the zero-field heat capacity data using formula $\Delta T_{ad} = -\int_0^H \frac{T}{C_p} \left(\frac{\partial M}{\partial T}\right)_H dH$. $S(H,T)$ can be evaluated by subtracting the corresponding $\Delta S_m$ from $S(0, T)$, where the total entropy $S(0,T)$ in absence of magnetic field is given by $S(0,T) = \int_0^T \frac{C_p(0,T)}{T} dT$. The temperature dependence of $\Delta T_{ad}$ for various magnetic fields is shown in Fig. 3(b). The maximum value of $\Delta T_{ad}$, increases nonmonotonically with increasing field and reaches as high as 21 K for a field change of 7 T (inset Fig. 3(b)). Besides large $\Delta S_m$ and $\Delta T_{ad}$ values, the refrigerant capacity (RC), which is a measure of the amount of heat transfer between the cold and hot reservoirs in an ideal refrigeration cycle, is an important quality factor for practical consideration of the magnetic refrigerant material. It is defined as $RC = \int_{T_1}^{T_2} \Delta S_m dT$, where $T_1$ and $T_2$ are the temperatures corresponding to both sides of the half-maximum value of $\Delta S_m$ (T) peak. Inset of Fig. 3 (a) shows the variation of RC with magnetic field. The maximum value of RC for a field change of 7 T is as large as 500 J kg$^{-1}$. Alternatively, the refrigeration cooling power (RCP), which is defined as $RCP = -\Delta S_m(\text{peak}) \times \delta T_{FWHM}$, where $\Delta S_m$(peak) is the peak value of the magnetic entropy change for a given field and $\delta T_{FWHM}$ is the full temperature width corresponding to half of the $\Delta S_m$(peak) is 643 J kg$^{-1}$.

To compare the magnetocaloric properties of single and polycrystalline samples, we measured the heat capacity of polycrystalline EuTiO$_3$ sample as a function of temperature from 100 K to 350 mK under $\mu_0 H$ = 0, 2, 5 and 7 T. For clarity, we show only the low-temperature data as $C_p/R$ where R is the gas constant in inset of Fig 4 (a). The zero-field data shows a clear peak at T = 5.4 K at the Neel temperature, which is slightly lower than that of the single crystal ($T_N$ = 5.6 K). The heat capacity peak is suppressed in magnitude, smeared and shifted towards higher temperature with increasing strength of the external magnetic field. The main panel of Fig.4(a) shows the normalized magnetic entropy ($S_m/R$) calculated from the heat capacity data under different magnetic fields. The theoretical magnetic entropy of Eu$^{2+}$ (S = 7/2) when spins are completely random is $S_m/R = \ln(2S+1) = \ln(8) = 2.079$. It can be seen from figure 4(a) that this temperature independent theoretical value is reached above 8 K. The zero-field magnetic entropy drops rapidly below $T_N$ and the value of the magnetic entropy is lowered under different magnetic fields as expected.

The magnetic entropy change ($\Delta S_m$) and the adiabatic temperature change ($\Delta T_{ad}$) for polycrystalline sample are shown in Fig. 4 (b) and (c), respectively, for field changes of $\mu_0 \Delta H =$ 2, 5 and 7 T. The $-\Delta S_m$ and $\Delta T_{ad}$ values are 47.32 J kg$^{-1}$K$^{-1}$ and 20 K, respectively for a field change of 7 T ($-\Delta S_m = 40$ J kg$^{-1}$K$^{-1}$ and $\Delta T_{ad} = 16.6$ K for $\mu_0 \Delta H = 5$T), which are slightly lower than the values found for the single crystal. The adiabatic temperature change observed in EuTiO$_3$ is larger than other promising magnetocaloric materials for refrigeration from 1 K to 20 K such as ErAl$_2$ ($\Delta T_{ad} = 12$ K at 13 K),[17] GdPd$_2$Si ($\Delta T_{ad} = 8$ K at $T = 17$K)[18] for $\mu_0 \Delta H = 5$ T and Dy$_3$Ga$_5$O$_{12}$ ($\Delta T_{ad} = 16$ K at $T = 2$ K)[19] for $\mu_0 \Delta H = 7$ T. On the other hand, the metal-organic framework material Gd(HCOO)$_3$ shows a comparable value of $\Delta T_{ad} = 22$ K for $\mu_0 \Delta H = 7$ T at 2 K.[20] The maximum value $-\Delta S_m = 42.4$ J kg$^{-1}$K$^{-1}$ in single crystal EuTiO$_3$ for a field change of 5 T is slightly larger than the maximum observed value 40 J kg$^{-1}$K$^{-1}$ for $x = 0.1$ in Eu$_{1-x}$Ba$_x$TiO$_3$ series,[13] and much larger than the highest value ($-\Delta S_m = 16$ J kg$^{-1}$K$^{-1}$ for $\mu_0 \Delta H = 5$ T for $R =$ Dy) found among other earth titanates ($R$TiO$_3$, $R =$ Dy, Ho, Er, Tm, Yb)[12]. The $-\Delta S_m$ in EuTiO$_3$ is larger than other Eu based compounds such as EuSe ($-\Delta S_m = 37.5$ J kg$^{-1}$K$^{-1}$ at $T = 4.6$ K),[21] Eu$_3$O$_4$ ($-\Delta S_m = 12.7$ J kg$^{-1}$K$^{-1}$ at $T = 5.3$ K),[22] and EuHo$_2$O$_4$ ($-\Delta S_m = 22.5$ J kg$^{-1}$K$^{-1}$ at $T = 5$ K)[23] for $\mu_0 \Delta H = 5$ T.

## IV. Summary

In summary, we have studied the magnetocaloric properties of single and polycrystalline EuTiO$_3$ samples using magnetization and heat capacity measurements. A large adiabatic temperature change of 21 K, isothermal entropy change of 49 J kg$^{-1}$K$^{-1}$, and refrigeration capacity of 500 Jkg$^{-1}$ for a field change of 7 T were found around the Neel temperature. This compound also show remarkable magnetocaloric effect even in low magnetic fields ($\Delta T_{ad}$ = 5 K, $-\Delta S_m$ = 9 J kg$^{-1}$K$^{-1}$ and $RC$ = 33 J kg$^{-1}$ for $\mu_0 \Delta H$ = 1 T). Our results suggest that the EuTiO$_3$ could be a potential material for cryogenic magnetic refrigeration below 20 K.

**Acknowledgements:**

R. M. thanks the Ministry of Education, Singapore for supporting this work through the grant no. R144-000-308-112. J.-S. Wang acknowledges the support of MOE Tier 2 grant R144-000-349-112. M. E. acknowledges financial support from MINECO through grant FEDER-MAT2012-38318-C03-01.

**Figure Captions:**

Fig. 1 (a) Left panel: Temperature dependence of dc magnetic susceptibility ($\chi$) under field-cooled (FC) and zero-field cooled (ZFC) processes. Right panel: Inverse susceptibility versus temperature for EuTiO$_3$ single crystal, where symbol and solid line represent the experimental data and Curie-Weiss fit. (b) Temperature dependence of zero-field heat capacity. Upper and lower insets show the heat capacity curve around $T_N$ and $T_{st}$, respectively.

Fig. 2 (a) Isothermal magnetization curves for EuTiO$_3$ single crystal. Inset: The five-segment magnetization curve at 2 K. (b) Experimental M-H curve (symbol) and theoretical fit using the equation (13) in Ref. 16.

Fig. 3 (a) Temperature dependence of isothermal magnetic entropy change ($-\Delta S_m$) for EuTiO$_3$ single crystal. Symbols: $-\Delta S_m$ calculated from isotherm magnetization data and solid lines: $-\Delta S_m$ calculated from mean field theory. Inset shows the refrigeration capacity (RC) as a function of magnetic field. (b) Temperature dependence of adiabatic temperature change ($\Delta T_{ad}$) for EuTiO$_3$ single crystal. Inset shows the maximum value of $\Delta T_{ad}$ as a function of magnetic field.

Fig. 4 (a) Main panel: Temperature dependence of magnetic entropy ($S_m$) of polycrystalline EuTiO$_3$ sample calculated from the heat capacity data. Inset shows the temperature dependence of heat capacity under different fields, (b) Magnetic entropy change ($-\Delta S_m$) and (c) adiabatic temperature change ($\Delta T_{ad}$) for the polycrystalline EuTiO$_3$ sample.

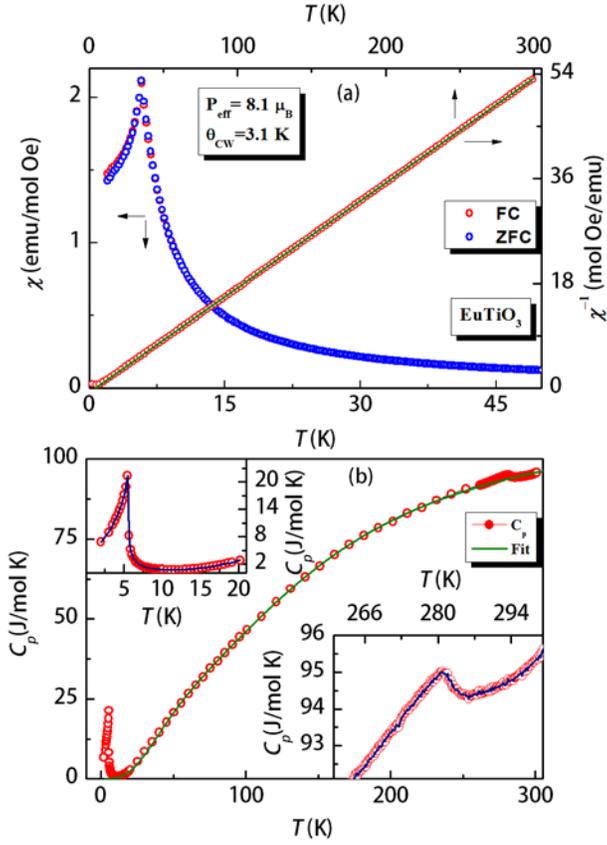

Fig. 1 (a) Left panel: Temperature dependence of dc magnetic susceptibility ($\chi$) under field-cooled (FC) and zero-field cooled (ZFC) processes. Right panel: Inverse susceptibility versus temperature for EuTiO$_3$ single crystal, where symbol and solid line represent the experimental data and Curie-Weiss fit. (b) Temperature dependence of zero-field heat capacity. Upper and lower insets show the heat capacity curve around $T_N$ and $T_{st}$, respectively.

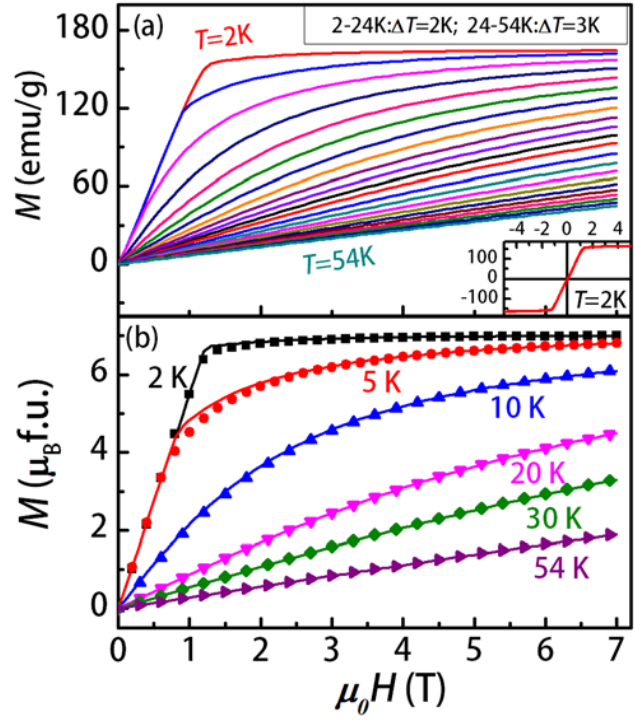

Fig. 2 (a) Isothermal magnetization curves for EuTiO$_3$ single crystal. Inset: The five-segment magnetization curve at 2 K. (b) Experimental M-H curve (symbol) and theoretical fit (solid line) using the equation (13) in Ref. 16.

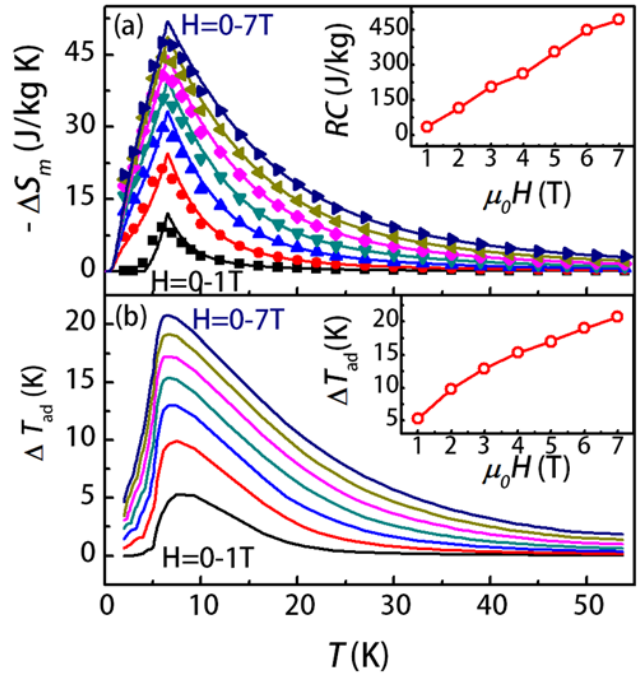

Fig. 3 (a) Temperature dependence of isothermal magnetic entropy change ($-\Delta S_m$) for EuTiO$_3$ single crystal. Symbols: $-\Delta S_m$ calculated from isotherm magnetization data and solid lines: $-\Delta S_m$ calculated from mean field theory. Inset shows the refrigeration capacity (*RC*) as a function of magnetic field. (b) Temperature dependence of adiabatic temperature change ($\Delta T_{ad}$) for EuTiO$_3$ single crystal. Inset shows the maximum value of $\Delta T_{ad}$ as a function of magnetic field.

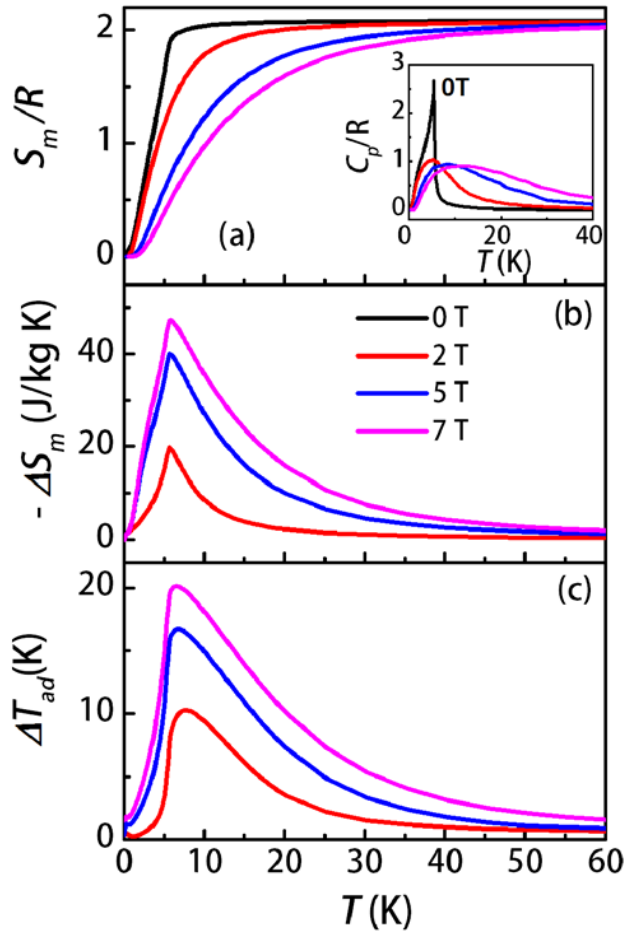

Fig. 4 (a) Main panel: Temperature dependence of magnetic entropy ($S_m$) of polycrystalline EuTiO$_3$ sample calculated from the heat capacity data. Inset shows the temperature dependence of heat capacity under different fields, (b) Magnetic entropy change ($-\Delta S_m$) and (c) adiabatic temperature change ($\Delta T_{ad}$) for the polycrystalline EuTiO$_3$ sample.